\documentclass[traditabstract]{aa}
\usepackage{graphicx}
\usepackage[dvips,pdf]{xcolor}
\usepackage{amssymb}
\usepackage{amsmath} 
\usepackage[varg]{txfonts}
\usepackage{booktabs}
\usepackage{nicefrac}
\usepackage{natbib}
\usepackage{float}
\usepackage{setspace}
\usepackage{hyperref}
\definecolor{darkgreen}{rgb}{0.2,0.6,0.}

\newcommand{\captionabove}[1]{\caption{#1}}
\bibpunct{(}{)}{;}{a}{}{,} 
\graphicspath{{figures/}}
\date{}
\title{The central star of the planetary nebula PB\,8:
       a Wolf-Rayet-type wind of an unusual WN/WC chemical composition %
 \thanks{%
     This
     paper includes data gathered with the 6.5-m Magellan Telescopes
     located at Las Campanas Observatory, Chile.
 }
 \thanks{%
     Some of the data presented in this paper were obtained from the
     Multimission Archive at the Space Telescope Science Institute (MAST).
     STScI is operated by the AURA, Inc.,
     under NASA contract NAS5-26555. Support for MAST for non-HST data is
     provided mainly by the NASA Office of Space Science via grant
     NAG5-7584.
     Based on INES data from the IUE satellite.
 }
}
\author{H. Todt\inst{1}      \and M. Pe\~{n}a\inst{2}   \and
        W.-R. Hamann\inst{1} \and G. Gr\"afener\inst{3}}
\institute{University of Potsdam, Institute for Physics and Astronomy,
           14476 Potsdam, Germany 
           \and Instituto de Astronom\'ia, Universidad Nacional
           Aut\'onoma de M\'exico, Apdo. Postal 70264, M\'exico
           D.F. 04510, M\'exico 
           \and Armagh Observatory, College Hill, Armagh BT61 9DG,
           Northern Ireland}  
\titlerunning{The central star of the planetary nebula PB\,8}

\abstract{%
  A considerable fraction
  of the central stars of planetary nebul\ae\  (CSPNe) are
  hydrogen-deficient. 
  As a rule, these CSPNe exhibit a chemical composition
  of helium, carbon, and oxygen with the majority
  showing Wolf-Rayet-like emission line spectra. These stars
  are classified as CSPNe of a spectral type [WC].
  We perform a spectral analysis of CSPN PB\,8 with
  the Potsdam Wolf-Rayet (PoWR) models for expanding
  atmospheres. The source PB\,8 displays wind-broadened emission lines
  from strong mass loss. Most strikingly, we find that its
  surface composition
  is hydrogen-deficient, but not carbon-rich. 
  With mass fractions
  of 55\% helium, 40\% hydrogen, 1.3\% carbon, 2\% nitrogen, and 1.3\%
  oxygen, it differs greatly from the 30--50\% of carbon
  which are typically seen in [WC]-type central stars. 
  The atmospheric mixture in
  PB\,8 has an analogy in the WN/WC transition type among the massive
  Wolf-Rayet stars. Therefore we suggest to introduce a new spectral
  type [WN/WC] for CSPNe, with PB\,8 as its first member. 
  The central star of PB\,8 has a relatively low temperature of
  $T_*=52\,\text{kK}$,
  as expected for central stars in their early
  evolutionary stages. Its surrounding 
  nebula is less than 3000 years old, i.e.\ relatively young. Existing
  calculations for the post-AGB evolution can produce
  hydrogen-deficient stars of the [WC] type, but do not predict the
  composition  found in PB\,8. We discuss various scenarios that
  might explain the origin of this unique object.
 } 
\begin{document}

\maketitle
\section{Introduction}

A planetary nebula (PN) surrounds a central star which is hot
enough ($T_\ast > 25\,000\,\text{K}$) to ionize its circumstellar matter.
According to the well-established scenario
{\citep[e.g.][]{paczynski1970,schoenberner1989}}, the central star of the 
planetary nebula (CSPN) ejects its nebula while suffering
thermal pulses at the tip of the asymptotic giant branch
(AGB). In the
subsequent PN phase it evolves rapidly towards the white
dwarf cooling sequence.

Most of the CSPNe show a hydrogen-rich surface composition.
Among the Galactic central stars, 5-6\,\%  are
hydrogen-deficient and show emission lines in
their spectra \citep{tylack1993,acknein2003}.
Moreover, half of these hydrogen-deficient stars
have spectra similar to 
those of massive Wolf-Rayet stars of the carbon sequence
and are therefore classified as spectral type [WC].
They have a strong stellar wind composed of helium, carbon, and
oxygen. Typical carbon surface-abundances have been found to lie
between 30\% and 50\% by mass \citep[see the reviews by][]{koesterke2001,crow2008}.
 
The central star (CS) of the planetary nebula PB\,8 (PN\,G292.4+04.1)
was first classified 
by \cite{mendez1991} as a hydrogen-rich Of-WR(H) star due
to the H$\gamma$ P Cygni profile and the appearance of an unusually
strong He{\,\sc ii} 4686 emission line.

In contrast, \cite{acknein2003} classified this star as a [WC5-6] type star.  

Below we analyze optical, IUE, and FUSE spectra of the central star
PB\,8 by means of the
Potsdam Wolf-Rayet (PoWR) model atmosphere code. The observations are
introduced in Sect.\,\ref{sect:pb8-obs}. 
Spectral modeling is briefly explained in Sect.~\ref{sect:pb8-methods}.
In Sect.~\ref{sect:pb8-analysis} we
describe the spectral analysis, and the results are discussed in the
final section (Sect.\,\ref{sect:pb8-discussion}). 

\section{Observations}
\label{sect:pb8-obs} 
\subsection{Optical spectrum}

High-resolution spectroscopy of PB\,8 was performed  on 2006 May 9 at Las
Campanas Observatory (Carnegie Institution) with the Clay 6.5m-telescope
and the double \'{e}chelle spectrograph MIKE (Magellan Inamori
Kyocera Echelle). This spectrograph operates with two
arms, which allow the observer to obtain blue and red spectra
simultaneously. The 
standard grating settings provided  wavelength coverage of
$3350-5050\,$\AA\  for the blue and $4950-9400\,$\AA\  for the
red. Three spectra with exposure times of 300\,s, 600\,s and 900\,s were
obtained. The slit width was $1\arcsec$ and was centered on the central
star. A binning of $2\times 2\,$pixels was used, providing a plate scale of
$0.26\,\arcsec$ per pixel. The spectral resolution varied from
$0.14\,$\AA\  to
$0.17\,$\AA\ in the blue and from $0.23\,$\AA\  to $0.27\,$\AA\  in the red as
measured with the comparison lamp.  

The data were reduced with standard procedures from the IRAF reduction
packages\footnote{IRAF is distributed by NOAO, which is operated by
 AURA, Inc., under contract to the National Science Foundation.}. 
Spectra were extracted with a $1.52\arcsec$ wide window and
flux-calibrated with respect to standard stars.
The three spectra were then weighted by exposure time and were finally
combined.

\begin{figure}[t]
{ 
\centering
\includegraphics[width=0.3\textwidth]{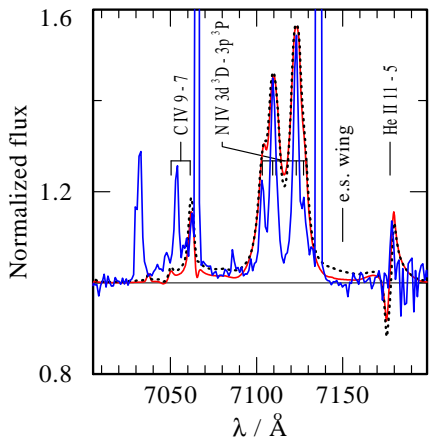}
\caption{
Electron scattering (e.s.) wings  for two similar models with same
$R_\text{t}$ but different density contrast $D$. The homogeneous model
(black dotted line) predicts stronger e.s. wings than observed (blue
solid). Despite the uncertainty due to normalization to the continuum,
the model with $D=10$ (red solid) seems to be more consistent with
the observation, although a higher clumping factor $D$ cannot be
excluded. The observation is rebinned to 1\,\AA\ for noise reduction.
\label{fig:pb8-clumping-NIV7100}
 }
}
\end{figure}

\begin{figure*}[!htb]
\centering
\includegraphics[width=\textwidth]{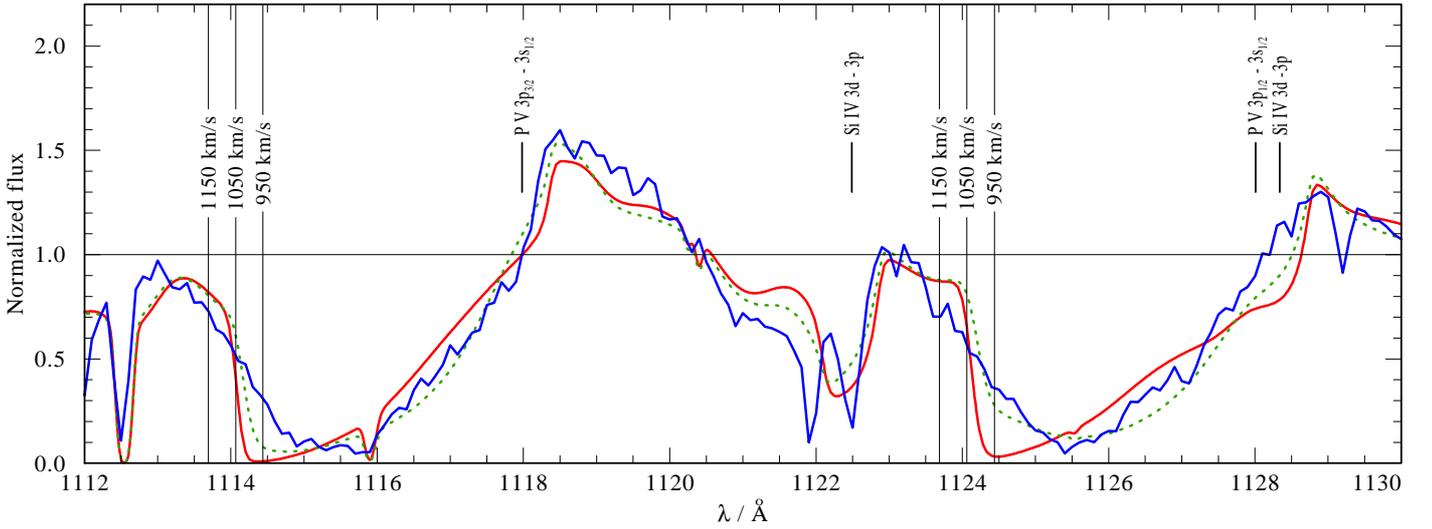}
\caption{Detail of the FUSE spectrum showing the P\,{\sc v} resonance
doublet, observation (blue line) vs. the final PoWR model (red solid
line), including ISM absorption. The round shape of the absorption
troughs is slightly better reproduced (green dashed line) when a
double-$\beta$ law is adopted for the velocity field (see text).
\label{fig:pb8-vinf-FUSE}
}
\end{figure*}

\subsection{UV spectra}
 
A low-resolution UV spectrum (1200 to 2000\,\AA), taken with the
International Ultraviolet Explorer (IUE) and a high-resolution FUV
spectrum (960 to 1190\,\AA), taken with the Far Ultraviolet
Spectroscopic Explorer (FUSE), were retrieved from the MAST
archive. For the UV range we used an exposure-time weighted
combination of the IUE spectra SWP28434LL 
($t_\text{exp} =2400\,\text{s}$) and  SWP30476LL
($t_\text{exp}=7800\,\text{s}$), both taken with the ``large''
IUE aperture. The spectral resolution was about 5\,\AA, and the
estimated S/N ratio is roughly 10 for $1200-1700\,$\AA\  and 20 for
$1700-2000\,$\AA. 

The FUSE observation of PB\,8 was performed with the LWRS aperture
of $30\arcsec \times 30\arcsec$ in run Z9111301000. We used a coadded
spectrum  ``all4ttagfcal'' from the CalFUSE pipeline, which
is already rebinned to $0.1\,$\AA\  to improve the S/N ratio, but is
still sufficient to resolve interstellar H$_2$ absorption lines.

\section{Methods}
\label{sect:pb8-methods}
\subsection{Spectral modeling}

For the spectral analysis we employed the PoWR models of expanding 
atmospheres. The PoWR code solves the non-LTE radiative transfer in a
spherically expanding atmosphere simultaneously with the statistical
equilibrium equations and accounts at the same time
for energy
conservation. Iron-group line
blanketing is treated by 
means of the superlevel approach \citep{graefener2002}, and wind
clumping in first-order approximation is taken into account
\citep{hamgrae2004}. 
We do not calculate hydrodynamically consistent models, but assume a
velocity field following a $\beta$-law with
$\beta=1$. Our present computations  include complex atomic models for
hydrogen, helium, carbon, oxygen, nitrogen, phosphorus, silicon, and
the iron-group elements.\\  
\\
After the computation of the synthetic spectrum, the models need to be
corrected for interstellar extinction. Dust extinction was taken into
account by the reddening law of \cite{cardelli1989}. Interstellar line
absorption in the FUSE range was calculated with the templates from
\cite{mccandliss2003} for H$_2$, and \cite{groenlam1989} for the Lyman series. 

\subsection{Spectral fitting}

The typical emission-line spectra of Wolf-Rayet stars are 
predominantly formed by recombination processes in their 
dense stellar winds. Therefore the continuum-normalized spectrum
shows a useful scale-invariance: for a given stellar temperature
$T_\ast$ and chemical composition, the equivalent widths of the
emission  lines depend in first approximation only on the ratio
between the volume emission measure of the wind  and the area of the stellar
surface. An equivalent quantity, which has been introduced by
\cite{schmutz1989}, is the {\em transformed radius} 
\begin{equation}
 R_\text{t} =
 R_\ast\left[\left. \frac{v_\infty}{2500\,\text{km}\,\text{s}^{-1}}
 \right/ \frac{\dot{M}\sqrt{D}}{10^{-4}\,\text{M}_\odot
 \,\text{a}^{-1}} \right]^{2/3}~~.  
 \label{eq:pb8-transradius}
\end{equation}
Different combinations of stellar radii $R_\ast$ and mass-loss 
rates $\dot{M}$ can thus lead to the same emission-line strengths. 
In the form given here, the invariance also includes
the micro-clumping parameter $D$, which is defined as the density
contrast between wind clumps and a smooth wind of the same mass-loss
rate. Consequently, mass-loss rates derived empirically
from fitting the emission-line spectrum depend on the adopted value
of $D$.  
The latter can be constrained by fitting the extended
electron scattering wings of strong emission lines
\citep[e.g.\ ][]{hamann1998}. 

\section{Analysis}

\label{sect:pb8-analysis}

Models for an un-clumped wind ($D = 1$) predict e.s.\ wings
to be stronger than observed. For the central star of PB\,8 we find that
$D=10$ is consistent with the observation
(cf. Fig.~\ref{fig:pb8-clumping-NIV7100}). 

For the terminal velocity $v_\infty$, we obtained a value of
$1000\,\text{km}\,\text{s}^{-1}$ from the width of the UV P-Cygni line 
profiles (cf. Fig.~\ref{fig:pb8-vinf-FUSE}). 
Note that the width of the absorption profile is matched by our
final model, but not the rather round shape of the profile. 
The observed shape of the absorption profile indicates a softer
increase of the velocity in the outer parts of the stellar wind, as
described, e.g., by a double-$\beta$ law
\citep{hilliermiller1999}.

A corresponding synthetic spectrum, where 40\,\% of the terminal
velocity are attributed to a second, flatter $\beta$-law with
$\beta_2=8$, is included in Fig.~\ref{fig:pb8-vinf-FUSE}.
From the limited agreement of the blue edges of the P\,{\sc v}
doublet, we estimate 
the uncertainty in $v_\infty$ to 
$\pm 100\,\text{km}\,\text{s}^{-1}$.

Line broadening by microturbulence is also included in
our models. From the shape of the line profiles we deduce a
microturbulence velocity of less than $50\,\text{km}\,\text{s}^{-1}$.

Stellar luminosity and mass were set to typical values for CSPNe, $L=
6000\,L_\odot$ and $M=0.6\,M_\odot$  
\citep[see e.g.][]{schoenberner2005,miller-bertolami2007}. 
The absolute flux of the model is diluted by its distance, 
which we consider to be a free parameter, as no certain distance is
known. With the help of Eq.~(\ref{eq:pb8-transradius}) the results can be
easily scaled to a different luminosity. The value of $M$ has no
noticeable influence on the synthetic spectra.

As aforementioned, the strength of WR emission lines mainly depends on 
the transformed radius $R_\text{t}$, the stellar
temperature $T_*$, and the chemical abundances.

We started our analysis by determining $R_\text{t}$ and
$T_*$ from a grid of models and then measured the abundances while the other
two parameters were kept fixed.

\begin{table}[t]
\begin{center}
\captionabove{
 Ratios between the peak heights: measured ratios, derived temperatures, 
 and ratios from the final model (model B). 
\label{tab:pb8-peakratios}
}
\begin{tabular}{llll}
\toprule
  & C\,{\sc iv} / C\,{\sc iii} 
& N\,{\sc iv} / N\,{\sc iii}  
& He\,{\sc ii}  / He\,{\sc i} \\
& 5800 / 4650 & 7100 / 4634 & 4686 / 5876 \\
\midrule
Observed & $1.4 \pm 0.3$  & $1.0 \pm 0.2$  & $2.3\pm 0.5$ \\
Derived $T_*$    & $50  \pm4\,$kK & $54 \pm 2\,$kK & $52\pm 3\,$kK \\
Final model   & 1.6            & 0.9            & 2.7  \\        
\bottomrule
\end{tabular}
\end{center}
\end{table}

Synthetic spectra from the grid of WNL model atmospheres\footnote{%
\href{http://www.astro.physik.uni-potsdam.de/~wrh/PoWR/powrgrid1.html}
{\tt http://www.astro.physik.uni-potsdam.de}} \citep{hamgrae2004}
were compared with the observed spectrum of PB\,8, giving a
first estimate
of $\log T_*\approx 4.7$ and $\log R_\text{t}/R_\odot = 1.4$.
From test calculations we obtained a first estimate of the chemical
composition (see Table~\ref{tab:PoWRmodelsABC}, model A). 
With these chemical abundances, an adopted luminosity
of $L=6000\,L_\odot$,
and a mass of $M=0.6\,M_\odot$ we computed a refined grid of models
around our first estimate $(T_*,R_\text{t})$.

\begin{figure}[t]
\centering
\includegraphics[angle=270,width=.45\textwidth]{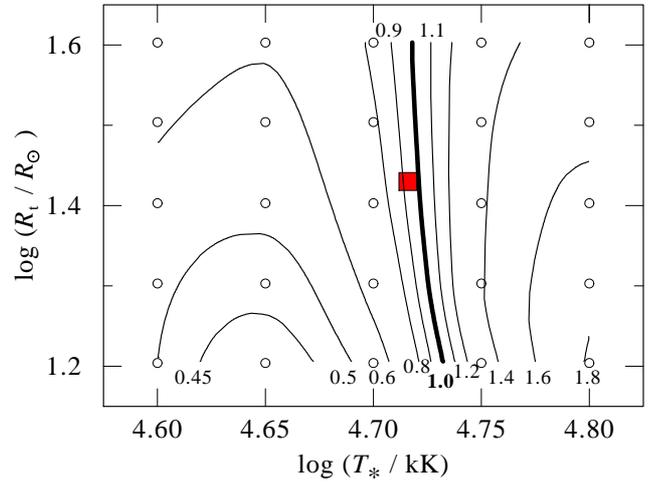}
\caption{
Contours of the ratio between the peak heights
of N\,{\sc iv} 7100 to N\,{\sc iii} 4643.
The thick contour represents the measured value. 
The open circles indicate the calculated models. Between these data
points the contour lines are interpolated. The best-fitting model
for PB\,8 is indicated by the red square.
\label{fig:iso_peak_NIV_7100-NIII_4643}
}
\end{figure}

\begin{figure*}[!tp]\begin{center}
\includegraphics[width=\textwidth]{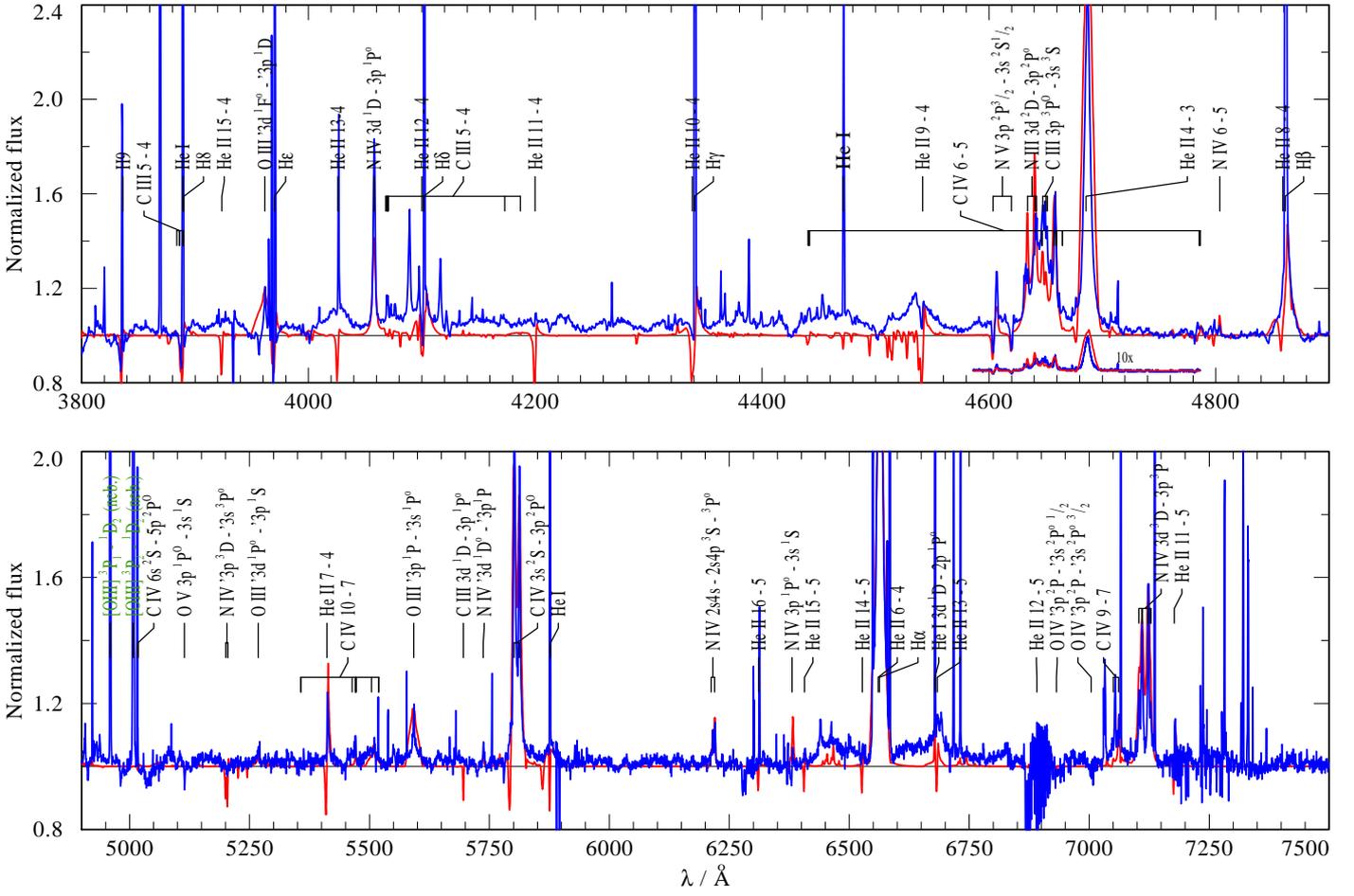}
\caption{Optical spectrum: observation of PB\,8 (blue, thin line) and
         best-fitting
         PoWR model (red, thick line), both normalized to the model
         continuum. The observation is rebinned to $0.5\,$\AA\ . The
         observed stellar spectrum is contaminated by the narrow
         nebular lines.
         \label{fig:pb8-optical_panels}}
\end{center}
\end{figure*}

For this grid, the line-strength ratios C\,{\sc iv} 5800 / C\,{\sc iii} 4650, 
N\,{\sc iv} 7100 / N\,{\sc iii} 4634,  and He\,{\sc ii} 4686 / He\,{\sc i} 5876 
were calculated and plotted
as contour lines over the grid. The contour plot for N\,{\sc iv} 7100 /
N\,{\sc iii} 4634 is shown in
Fig.~\ref{fig:iso_peak_NIV_7100-NIII_4643}.
Using line ratios instead of the absolute line strengths diminishes the influence of 
chemical abundances. Moreover,
Fig.~\ref{fig:iso_peak_NIV_7100-NIII_4643} shows that in the parameter
range under consideration the line ratio depends almost only on the
temperature and not on the transformed radius. In this way,
the temperature determination de-couples from $R_\text{t}$.
The equivalent width of the C\,{\sc iii} 4650 line
cannot be measured accurately, because this line is
partly overlapping with the N\,{\sc iii} 4634 line 
(see Fig.~\ref{fig:pb8-optical_panels}).
Therefore we used the peak height of the strongest multiplet component 
as a measure of the line strength, which is
less affected by blending than the equivalent width. 
Peak ratios from observation and models are listed in Table~\ref{tab:pb8-peakratios}. 
From our optical observation we
estimate the uncertainty in the normalized continuum to be on the
order of $10\,\%$. We consider this to be the uncertainty in the peak
measurement, which we then use to infer a $20\,\%$ error in the measured
ratio. 

As included in Table~\ref{tab:pb8-peakratios}, slightly different
stellar temperatures are derived from the different elements.
But within the inferred uncertainties a temperature of
$T_*=52\,\text{kK}$ is consistent with all three observed line ratios. 

\begin{table*}[!ht]
{\begin{center} 
\captionabove{Parameters for the PoWR models as shown in
  Fig.~\ref{fig:pb8-optical_comparison}
and \ref{fig:pb8-optical_panels}.
Our final model is model B.
\label{tab:PoWRmodelsABC} }
\begin{tabular}{lrrrrrlllrr}
\toprule
model   & \multicolumn{1}{c}{$\log R_\text{t}$} & \multicolumn{1}{c}{$\log\dot{M}$}  & \multicolumn{1}{c}{$T_*$} & He & H  & C & N & O  & \multicolumn{1}{c}{$d$} & \multicolumn{1}{c}{$E_{\text{B}-\text{V}}$} \\
        & \multicolumn{1}{c}{$[R_\odot]$}  & \multicolumn{1}{c}{$[M_\odot\,\text{a}^{-1}]$} & \multicolumn{1}{c}{[kK]}    & \multicolumn{5}{c}{--- \% mass fraction ---} & \multicolumn{1}{c}{[kpc]} & \multicolumn{1}{c}{[mag]}\\ 
\midrule 
 A & 1.35         & $-6.89$ & 50    & 66 & 30 & 1   & 1.5 & 1   & 4.6 & 0.40\\
 B & 1.43         & $-7.07$ & 52    & 55 & 40 & 1.3 & 2   & 1.3 & 4.2 & 0.41\\
 C & 1.55         & $-7.20$ & 50    & 33 & 60 & 2   & 3   & 2   & 4.4 & 0.42\\
\bottomrule
\end{tabular}
\end{center}}
\end{table*}

\begin{figure*}[!htp]
\centering
\includegraphics[width=\textwidth]{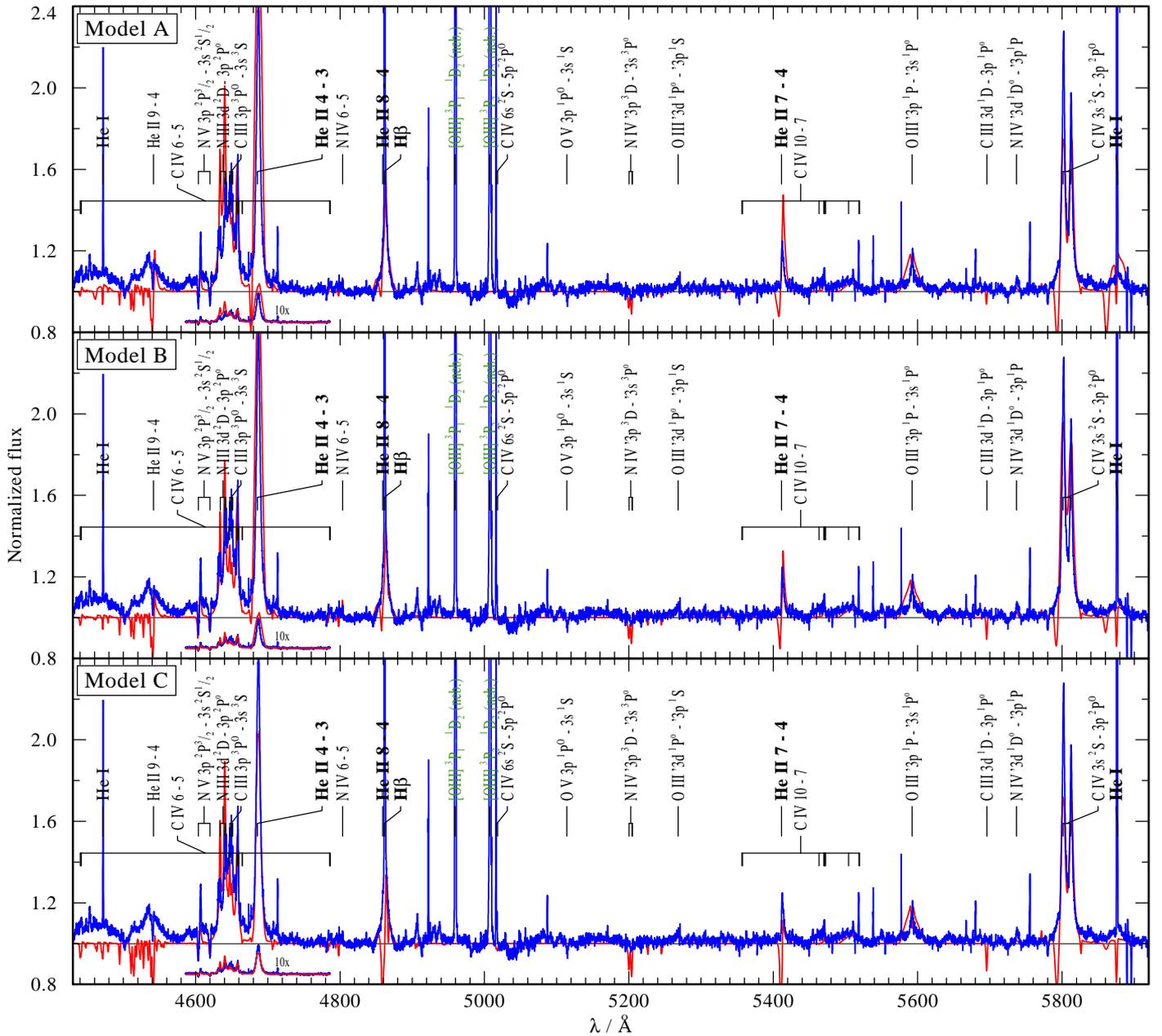}
\caption{PB\,8: Detail of the optical spectrum,  comparison of the
  PoWR models A, B, and C (red lines) vs. observation (blue). 
  PoWR model~A (upper panel) with only 30\,\% hydrogen. 
  The emission lines from helium appear much too strong.
  PoWR model~C (lower panel) with reduced mass-loss rate
  relative to model~A and reduced helium abundance. To recover the
  strength of the 
  emission lines of carbon, nitrogen, and oxygen of model~A, their
  abundances are 
  increased. The helium emission lines are weaker than
  observed. This model is considered as an upper limit for the
  hydrogen, carbon, oxygen, and nitrogen abundance, and as a lower
  limit for the mass-loss rate and the helium abundance.
  The compromise between models A and C and therefore our final model
  is model B, shown in the second panel.
\label{fig:pb8-optical_comparison}
}
\end{figure*}

Although the line strengths of most of the spectral lines of carbon,
oxygen, and nitrogen can be reproduced by our first estimate model A
(Fig.~\ref{fig:pb8-optical_comparison})
some of the unblended helium lines appear stronger than observed.

\begin{figure*}[!pt]
\includegraphics[width=\textwidth]{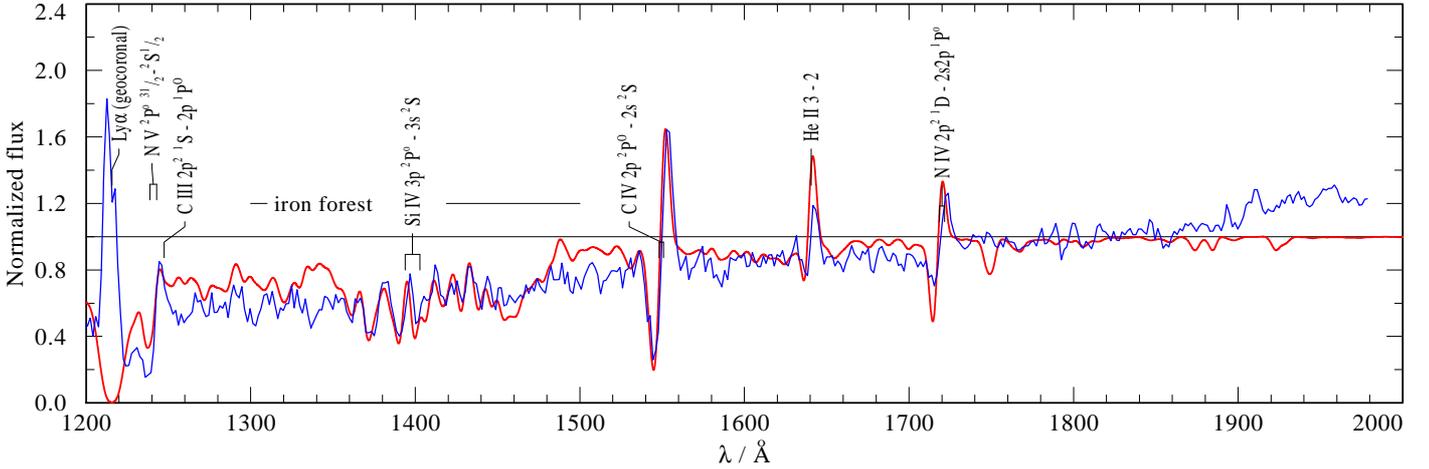}
\caption{Observed IUE spectrum of PB\,8 (blue, thin line) vs. PoWR
         model (red, thick). The observed spectrum was normalized by
         the model continuum. The synthetic spectrum was folded with a
         Gaussian with a FWHM of $5\,$\AA, corresponding to the spectral
         resolution of the IUE observation. 
         The iron forest is only partially reproduced.  
         \label{fig:pb8-IUE_panel}} 
\end{figure*}

\begin{figure*}[!tp]
\includegraphics[width=\textwidth]{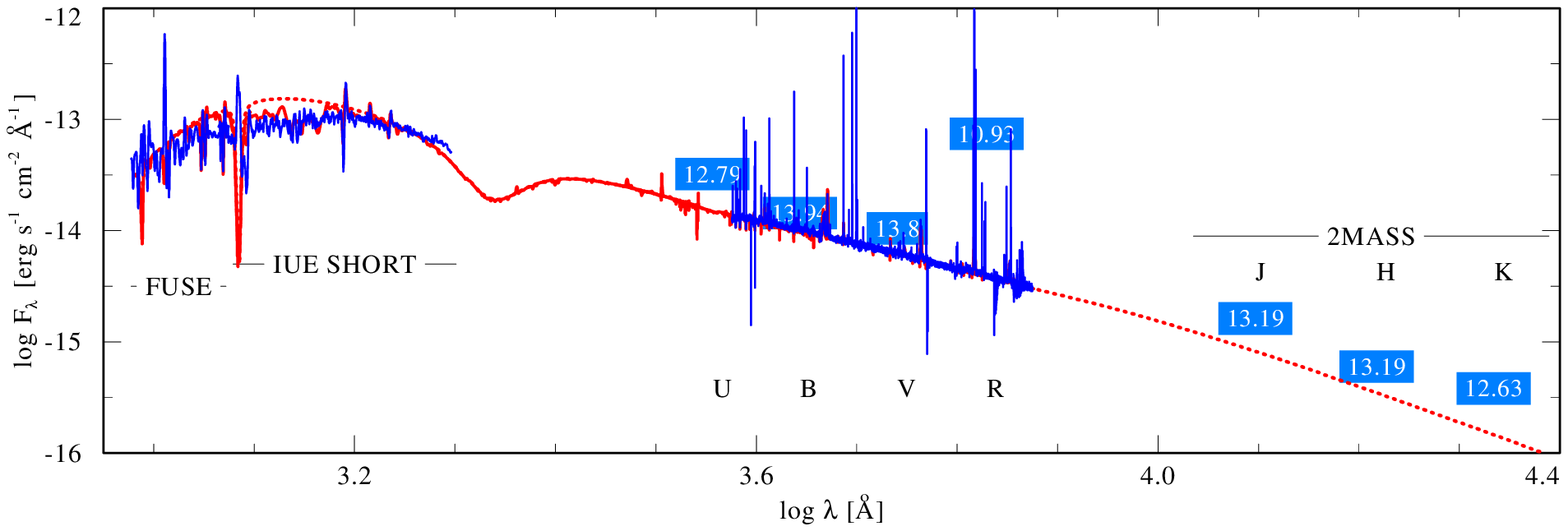}
\caption{Spectral energy distribution for CS PB\,8, model 
         vs.\ observation. Observed spectra (blue thin lines) are from
         FUSE, IUE, and MIKE (see Sect.\,\ref{sect:pb8-obs}). 
         Photometric values (blue blocks) taken from 
        \cite{acker1992} for UBV, \cite{gsc2001} for R, and 2MASS
        \citep{skrut2006} for JHK are partly contaminated by nebular
         emission.   
         The calculated spectrum (red line) is for the model
         parameters in Table~\ref{tab:pb8-parameters}. The model flux
         was reddened with $E_\text{B-V}=0.4$ and  $R_\text{V}=4$
         and corrected for interstellar Lyman line absorption. The
         model continuum without lines is also shown for comparison
         (red dotted). Note that in the IUE SHORT range the iron
         spectral lines form a pseudo-continuum. 
\label{fig:pb8-sed}} 
\end{figure*}

Therefore we calculated a model with the same $T_*$, but 
half the mass-loss rate ($2^{2/3}\times R_\text{t}$) and half
the helium-mass fraction (model C in Table~\ref{tab:PoWRmodelsABC}). To
recover the line strengths of H, N, O, and C, the mass fraction for
each of these elements is thus doubled. 

Then, as expected, all of the spectral lines of helium are weaker than
in model A (Fig.~\ref{fig:pb8-optical_comparison}). The He\,{\sc ii}
4686 line in model C however seems to be more consistent with the
observation. In contrast, 
the line blend of He\,{\sc ii} 4859 and H$\beta$ exhibits deep
absorption features, which are not observed. Furthermore, the modeled
He\,{\sc i} 5876 line is now much too weak in the model. 

For our final fit we thus chose a model with parameter values in
between those of model A and C.
As a compromise, the best fit of all 
spectral lines is achieved with our final model
B at a stellar temperature of $T_* = 
52\,\text{kK}$ and a transformed radius of $R_\text{t} =
26.9\,\text{R}_\odot$ (cf. Figs.~\ref{fig:pb8-optical_panels} and
\ref{fig:pb8-IUE_panel}). 

Parallel to the fitting of the normalized spectrum
we obtained the synthetic spectral energy distribution (SED) in
absolute units. This model-SED was fitted to the calibrated spectra 
and photometric measurements by adjusting the distance and the
reddening parameter $E_{\text{B} -\text{V}}$ 
(see Fig.~\ref{fig:pb8-sed}). 

\begin{figure}[!h]
\begin{center}
\resizebox{\hsize}{!}{\includegraphics{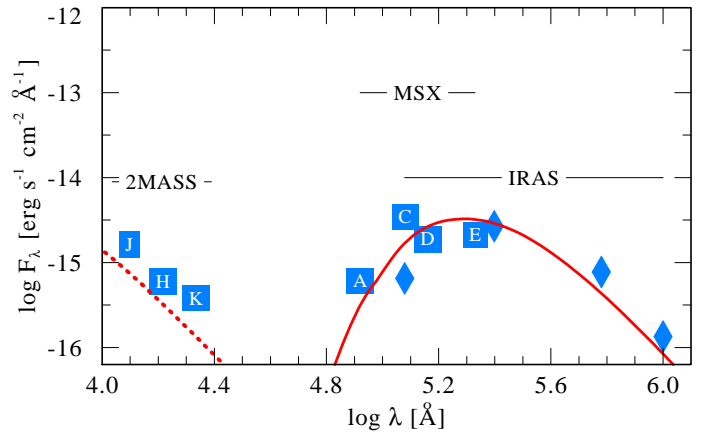}}
\caption{SED for PB\,8 in the infrared range. Photometric
         observations (blue blocks) for JHK are from 
         2MASS, ACDE from MSX (MSXPSC V2.3), and $12, 25, 60,
         100\,\mu\text{m}$ (blue diamonds) from  IRAS (IPAC V2.0
         1986). Also shown is 
         the synthetic stellar continuum (red dotted) and a  blackbody
         spectrum (red line) for $T=150\,\text{K}$, indicating
         warm dust emission. 
 \label{fig:pb8-iras}
}
\end{center}
\end{figure}

The best SED-fit was obtained with a color excess of $E_{\text{B}-\text{V}}
= 0.41$ and $R_\text{V}=4$ \citep{cardelli1989}.
This value is considerably higher than the value of
$E_{\text{B} -\text{V}} = 0.24$ derived for the nebula from the Balmer
line decrement by \cite{gar2009} for the same observations,
but with an adopted value of $R_\text{V} =3.1$. Following
\cite{cardelli1989}, a higher 
value of $R_\text{V}$ can be interpreted as a larger dust
grain size, meaning there may be different dust compositions along the
line of sight towards the planetary nebula and the central
star. 
A possible explanation could be dust within the nebula in the close vicinity of
the central star, as there is a strong mid-IR
emission visible in MSX and IRAS observations
(Fig.~\ref{fig:pb8-iras}). The comparison with a blackbody
with $T=150\,\text{K}$ indicates that this emission might be caused
by ``warm'' dust. The 2MASS photometry values are higher than
predicted by the stellar atmosphere model, presumably due to contamination by
nebular emission.

With the adopted stellar luminosity of $6000\,L_\odot$ we estimated
a distance of 4.2\,kpc towards PB\,8. This value is of the same order
of magnitude as the $5.15\,\text{kpc}$ which were derived from 
the nebular luminosity and its brightness temperature in the radio
range (5\,GHz) and nebular distances, which lie between 2.2 and 5.8\,kpc 
\citep[][and references therein]{phillips2004}.

\begin{figure*}[!tp]
\includegraphics[width=\textwidth]{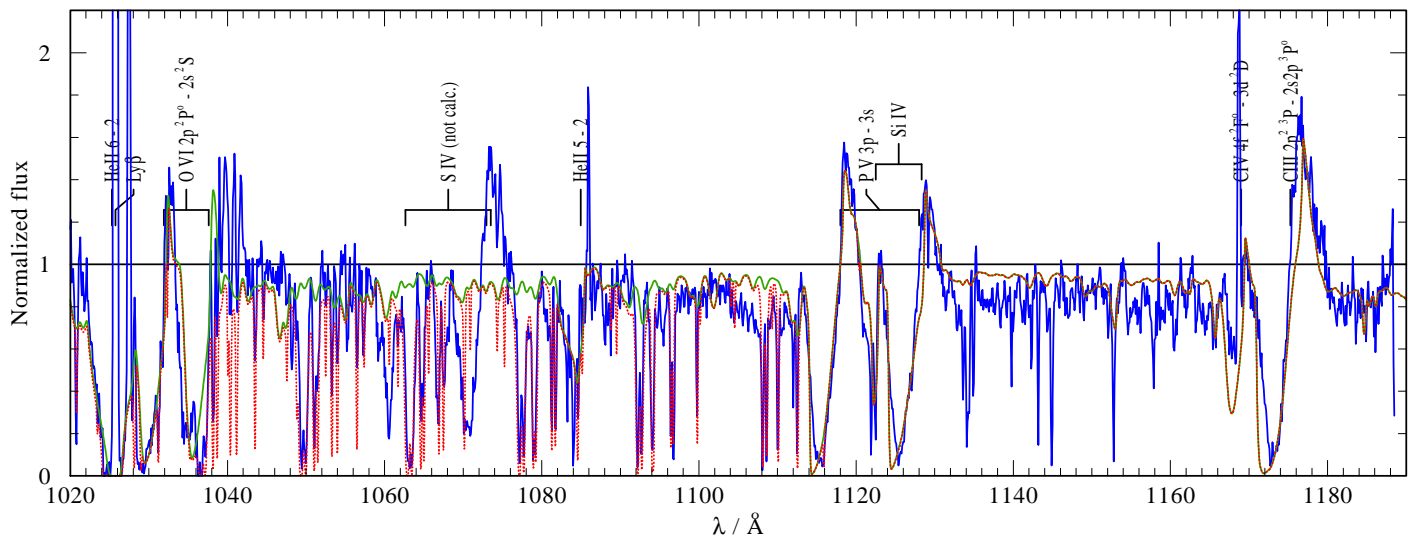}
\caption{Detail of the normalized FUSE spectrum (blue line) 
         compared to the synthetic spectrum after accounting for
         interstellar H$_2$ and Lyman absorption (red dotted). The
         stellar spectrum shows strong P Cygni profiles. Superimposed
         are narrow interstellar absorption lines of H$_2 $, emission
         lines from the planetary nebula, and telluric air glow
         features. The synthetic spectrum without interstellar H$_2$
         absorptions is shown for comparison (green).
  \label{fig:pb8-FUSE}}
\end{figure*}

The observed FUSE flux cannot be reproduced consistently with the IUE
and optical flux by any set of model parameters, distance or
extinction. In particular, the flux between $1140\,$\AA\  and
$1200\,$\AA\  seems to be too low compared with the model. This might
be caused by problems with the channel alignment, as reported for
other observations, e.g.\ \cite{miksa2002}. We checked the unbinned
FUSE spectra from the program Z911, target 13 and found a discrepancy
in the measured fluxes between the 2ALiF and 1BLiF channels of up to
50\% in the range of $1140\,$\AA\ to $1200\,$\AA. Therefore we
applied a manual correction to the rebinned spectrum in this
wavelength range (Fig.~\ref{fig:pb8-FUSE}) to match the overlapping IUE flux.

FUSE spectra are usually affected by H$_2$ absorption lines,
nebular and airglow emission lines, which hamper the analysis of the
underlying stellar spectrum. A simple fit of the H$_2$ absorption
lines by eye was performed with the help of adequate templates to tell
interstellar from stellar absorption lines. For the fitting we used a
column density of $n_{\text{H}_2}(J=0) = 4\times10^{19}
\,\text{cm}^{-2}$ and excitation temperatures of $T_{J=0-J=1} =
60\,\text{K}$ and $T_{J=2-J=4} = 270\,\text{K}$.  The synthetic
stellar spectrum, which was corrected for this interstellar absorption
from H$_2$, is shown in Fig.~\ref{fig:pb8-FUSE}. We found the same
radial velocity of $v_\text{rad}=12\,\text{km}\,\text{s}^{-1}$ for the
central star, its planetary nebula, and the absorbing ISM. 

To reproduce the observed P Cygni profiles of the O{\,\sc vi}
resonance doublet in the FUSE spectrum, 
super-ionization by X-ray emission was included in the model. 
For this purpose an optically thin hot gas component of $T=1.5\,$MK
was assumed to be distributed within the stellar wind. We only accounted
for its free-free emission (thermal bremsstrahlung), because the filling
factor was arbitrarily chosen \citep[cf.][for the formalism]{baum1992}.

From our PoWR model (without additional X-ray emission) we predict an
H{\,\sc i} Zanstra temperature of
$T(\text{H{\,\sc i}}) = 53 \pm 3\,\text{kK}$. \cite{shawkaler1989} measured
a lower value, $T(\text{H{\,\sc i}}) =
32\pm3\,\text{kK}$. This discrepancy could mean that the nebula is not
optically thick in all directions of the ionizing radiation. For
He{\,\sc ii} the model predicts a Zanstra temperature of only
$T(\text{He{\,\sc ii}}) = 21 \pm 1 \,\text{kK}$ due to the
strong absorption in the helium-rich wind, which means that there
should not be enough photons to create a noticeable zone of fully
ionized helium in the nebula. This agrees with \cite{gar2009}, who
could not detect any nebular He{\,\sc ii} lines in the spectrum of
PB\,8.  

\subsection{Element abundances}
\label{sect:pb8elements}
\begin{table}[h]
{\begin{center} 
\captionabove{Parameters of PB\,8
         \label{tab:pb8-parameters}}
\onehalfspacing
\begin{tabular}{lrl}
\toprule
$T_*$              & $52\pm 2$           & $\text{kK}$              \\
$v_\infty$         & $1000\pm 100$       & $\text{km}\,\text{s}^{-1}$      \\
$\log \dot{M}$          & $-7.07^{+0.17}_{-0.13}$ & $M_\odot\,\text{a}^{-1}$ \\
$\log R_\text{t}$       & $1.43_{-0.08}^{+0.12}$& $R_\odot $     \\
$E_{\text{B} - \text{V}}$    & $0.41 \pm 0.01$     & $\text{mag}$             \\
$d(L_*=6000\,
 \text{L}_\odot)$  & $4.2 \pm 0.2$       & $\text{kpc}$             \\
$v_\text{rad}$     & 12                  & $\text{km}\, \text{s}^{-1}$     \\
H                  & $40_{-10}^{+20}$    & \% mass fraction      \\
He                 & $55_{-22}^{+11}$    & \% mass fraction      \\
C                  & $1.3_{-0.3}^{+0.7}$ & \% mass fraction      \\
N                  & $2.0_{-0.5}^{+1.0}$ & \% mass fraction      \\
O                  & $1.3_{-0.3}^{+0.7}$ & \% mass fraction      \\
Fe                 & $1.6\times 10^{-3}$ & \% mass fraction      \\
P                  & $5.2\times 10^{-6}$ & \% mass fraction      \\
Si                 & $3.2\times 10^{-4}$ & \% mass fraction      \\
\bottomrule
\end{tabular}\\[.3cm]
\singlespacing
\end{center}
}
\end{table}

\begin{figure*}[!tp]
\begin{center}
\includegraphics[bb=130 510 464 662,clip=,height=7cm]{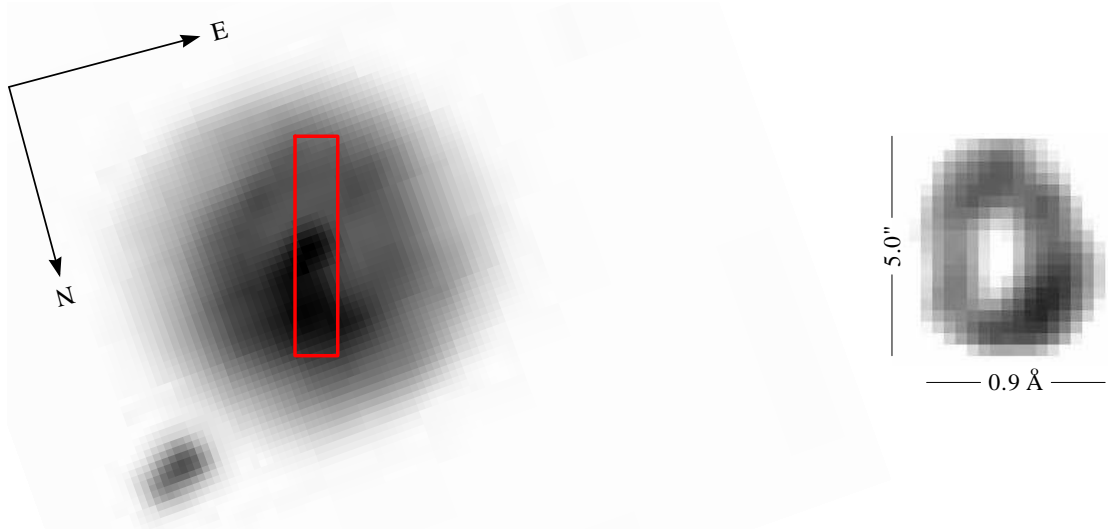}
\caption{{\bf Left:} Composite picture of PB\,8 from H$\alpha$, [N{\,\sc ii}], and
           [O{\,\sc iii}] emission lines with linear intensities
           \citep[adapted from][]{schwacorr1992}. 
           Also shown is the slit position (red rectangle) of our
           optical observation.
           The slit orientation was set to the parallactic angle, which
           actually varied between $-20^\circ$ and $-10^\circ$ during
           the exposure. 
           The bright nebular knot in the northern part of the slit can
           also be recognized in the lower part of the spectrogram
           (cf. right panel).
           \label{fig:pb8-linear}
           {\bf Right:}
           Spectrogram of the [O{\,\sc iii}] 5007\AA{} line, indicating
           the roughly spherical symmetry of the nebula. This symmetry
           is seen along the slit as well as in the dispersion
           direction, which reflects the radial velocity along the
           line-of-sight. The southern end of the slit is to the top
           (cf. left panel).
           }
\end{center}
\end{figure*}  

\emph{Hydrogen.}
He{\,\sc ii} lines from the Pickering series with even principle quantum
numbers $n$
appear much stronger in the observation than the odd members of the
same series. As the former are blended with the Balmer lines of hydrogen,
this is clear evidence for a significant contribution from hydrogen. We
obtained the best fit with a hydrogen mass fraction of $40\,\%$.
\\
\\
\noindent
\emph{Carbon.} The carbon mass fraction derived from C{\,\sc iii} and
C{\,\sc iv} lines is only $X_{\rm{C}} = 1.3\,\% $. Spectral lines from
other ionization stages, especially C{\,\sc ii}, are not detected.
\\
\\
\noindent
\emph{Oxygen.} The oxygen abundance is based on the emission lines
from O{\,\sc iii} and O{\,\sc iv}. The O{\,\sc vi} resonance line at
$1031\,$\AA\  in the FUSE spectrum depends on the super-ionization effect 
and is therefore not useful for the abundance determination.
\\
\\
\noindent
\emph{Nitrogen.} A nitrogen abundance of about $2\,\%$ by mass
is derived from the spectral lines of N{\,\sc iii} and N{\,\sc iv}. 
The N\,{\sc v} lines appear too weak in the model.  
Note that similar to O\,{\sc vi} the excitation of
N\,{\sc v} might be dominated by super-ionization.
\\
\\
\noindent
\emph{Iron.}
Iron is included with a slightly super-solar abundance 
($X_\text{Fe} = 1.6\times 10^{-3}$ by mass). For the iron-group elements 
Sc, Ti, V, Cr, Mn, Fe, Co, and Ni, a given relative
abundance with respect to iron is adopted, as outlined in
 \cite{graefener2002}.
The iron forest, visible in the FUSE 
and IUE range below $1500\,$\AA, is roughly reproduced with
these abundances. 
\\
\\
\noindent
\emph{Phosphorus and silicon.}
With solar abundances, the P\,{\sc v} and Si\,{\sc iv} lines in the
UV spectra are well fitted, see e.g.\ Fig.~\ref{fig:pb8-vinf-FUSE}.
\\
\\
Stellar parameters and chemical abundances derived for the central
star of PB\,8 are compiled in Table~\ref{tab:pb8-parameters}. 

Note that the errors given there are inferred from the sequence of our PoWR models 
A, B, C.

\subsection{Nebula age} \label{sect:pb8-nebula}

For the nebula, \cite{gar2009} derived an expansion velocity of
$14\pm 2\,\text{km}\,\text{s}^{-1}$ from the separation of the 
maxima of [O{\,\sc iii}] $\lambda\, 5007$. 
With the same method we found an
expansion velocity of $19\pm 6\,\text{km}\,\text{s}^{-1}$
from [N{\,\sc ii}] $\lambda\,6548$ and $6583$. 
Following \cite{schoenberner2005b},  this kind of discrepancy 
is characteristic
for young planetary nebul\ae, as the individual emission lines form in
different regions with different velocities in
the beginning of the PN expansion. 
Especially for the determination of the
kinematic age of the nebula, only the [N{\,\sc ii}] shell
 is a reliable
indicator \citep{schoenberner2005b}.
Therefore we estimated the shell radius from the [N{\,\sc ii}]
spectrogram, obtaining a value of $2.5\arcsec$. 
No substructure of the
[N{\,\sc ii}] doublet, which would indicate different velocities for rim and shell,
could be resolved. 
Thus we adopt the peak separation as the indicator for the shell expansion velocity
and regard this as a lower limit.
Together with the shell radius and the
spectroscopic distance, this yields an upper limit for the  dynamical age of the
nebula of {$2600\,$}years, which agrees with \cite{gesicki2006}. Hence
 we conclude that PB\,8 is a 
relatively young nebula.

\section{Discussion}
\label{sect:pb8-discussion}

\subsection{PN and Central star status}
\label{sect:pb8-PNstatus}

The nebula PB\,8 appears as a roughly spherical nebula,
nearly round in the composite image of H$\alpha$, 
[N{\,\sc ii}], and [O{\,\sc iii}]
(Fig.~\ref{fig:pb8-linear}, left panel
\footnote{from {\tt http://www.astro.washington.edu/balick/PNIC/}}), 
although the shell shows some knotty structure. In particular, there 
is a bright structure extending from the center to the northern side
of the shell. 
The long-slit spectrogram also reveals a good symmetry in the radial
velocities 
(Fig.~\ref{fig:pb8-linear}, right panel).
Given the unique chemical abundances of the central star PB\,8, one
must consider the possibility that this object is actually a massive
star with a ring nebula. However, the low nebular expansion velocity
discussed in Sect.~\ref{sect:pb8-nebula} is rather characteristic for
PNe. 
\cite{medina2006} found expansion velocities for PNe with
 Wolf-Rayet nuclei in the range of
 $8-44\,\text{km}\,\text{s}^{-1}$ from direct observations. 
Expansion velocities for ring
nebul\ae\  around massive stars are systematically higher,
$16-110\,\text{km}\,\text{s}^{-1}$ \citep{chu1999}.

Moreover, the electron density in PB\,8 measured by \cite{gar2009}, 
${n_\text{e} = 2550\pm 550\,\text{cm}^{-3}}$ is typical for young
planetary nebul\ae, but several times higher than found in ring
nebul\ae\ \citep{mathis1992}. 

Furthermore, if the central star of PB\,8 were a massive star, this
would imply a luminosity of at least $\log (L/L_\odot) = 5.3$,
which shifts the distance  to {$\approx 24.2 \,\rm kpc$. With a Galactic
latitude of $4^\circ$ this corresponds to a height of $1.7{\,\rm kpc}$}
above the fundamental plane of the Galaxy. This is much more than the
scale height of the thin disk and therefore an unlikely location for a
massive star. 

\subsection{Re-classification of the central star of PB\,8}

The central star of PB\,8 has been classified as spectral type [WC5-6]
by \cite{acknein2003}. Yet we showed above that the central
star of PB\,8 is not a member of the [WC] sequence; its spectrum shows
strong lines of nitrogen, reflecting that its chemical composition
rather resembles that of a WN star. Nevertheless, carbon is slightly
enhanced, in contrast to the typical WN composition where carbon is
strongly depleted due to the CNO cycle equilibrium.

Among massive WR stars are a few objects with a similar
composition as our program star, which are usually considered to be
caught in the transition phase between the WN and WC stage. These are
classified as spectral type WN/WC or WNC. Therefore in analogy to
these massive stars we suggest to classify the central star of PB\,8
as [WN/WC].   

The detailed subtype of PB\,8 is WN6 when
applying the classification scheme established by \cite{smith1968} for
massive WN stars. With the scheme in \cite{vanderhucht1981} for massive
WC stars, the WC7 subtype seems to be appropriate. In combination with
these two schemes, we determine the  detailed subtype classification
as [WN6/WC7]. 

The [WC5-6] classification of PB\,8 by \cite{acknein2003} was
partly based on the identification of spectral features with stellar
C\,{\sc ii}, but we cannot confirm any stellar C\,{\sc ii} line  from
our high-resolution data.

\cite{tylack1993} alternatively defined the class of ``weak emission
line stars'' (WELS) for those spectra that show much fainter and
narrower emission lines than massive WC stars. \cite{gesicki2006}
assign this WELS classification to PB\,8. However, the nature and
homogeneity of the WELS class seems to be still unclear.

There are two other known WR-type central stars with non-carbon-rich
winds. One is LMC-N\,66 in the LMC, which is only sometimes of the
Wolf-Rayet type. It has an irregular nebula and seems to be a close
binary \citep[see discussion in][]{pena2004}. The other example, the
central star of PMR\,5 discovered by \cite{morgpark2003}, is probably
a Galactic [WN] star. Its spectrum shows only helium and nitrogen
lines, while any carbon lines are missing. In case of PB\,8, carbon
and oxygen lines are visible. \cite{morgpark2003}
discuss the PN status of PMR\,5 on the basis of the nebular
expansion velocity and electron density. They conclude
that PMR\,5 is a normal PN. 

\subsection{Evolutionary status}

The surface composition of PB\,8 appears unique among all
CSPNe that have been analyzed so far. Only two
other CSPNe (PMR\,5 and the enigmatic variable LMC-N\,66) are known to
show a WN-type composition, which is dominated by helium
with a significant amount of nitrogen. Two more CSPNe are known to be
helium-rich, but without strong winds 
\citep[LoTr\,4 and K\,\mbox{1-27},][]{rauch1998}. 
Our program star PB\,8 is unique in
showing a significant amount of carbon, while carbon is usually
depleted in WN-type compositions. 

Note that there is a He-sdO star without PN, KS\,292, that shows a similar
composition as PB\,8, including the enhanced carbon abundance
\citep{rauch1991}.
 
This poses the question of how to
explain the evolutionary origin of PB\,8.

The formation of hydrogen-deficient post-AGB stars can be explained by a
final thermal pulse which leads to the ingestion of the hydrogen envelope
\citep{herwig1999,herwig2001,werner2006,althaus2005}.
This final thermal pulse may occur either at the tip of the AGB (AGB
final thermal pulse, AFTP) or later, when the AGB has been left (late
thermal pulse, LTP, or very late thermal pulse, VLTP). These models
lead to a carbon-rich surface composition (carbon abundance larger
than $20\,\%$ by mass), which is what is needed to explain the observed
abundance patterns of [WC]-type central
stars. However, the predicted nitrogen abundance
is very small, except for the VLTP case where $X_\text{N} \geq 1\%$
has been predicted \citep{althaus2005,werner2006}. 
 
As a tentative explanation for PB\,8, we propose that the final thermal
pulse has only been ``weak'', so that only a small amount of carbon
has been dredged up to the surface. The bulk of matter 
at the surface is then enriched by helium from the former
intershell region. This material also 
contains nitrogen according to the equilibrium from the CNO
cycle. In addition, a part of the hydrogen-rich envelope must have
survived the last pulse and become mixed into the present outer
layers. While it is not clear whether such a ``weak'' last thermal
pulse can happen on the AGB, it might occur in an extremely late VLTP
when the star is already too cool to undergo a full He-shell flash
(F.\,Herwig, private communication). 

Further constraints for the evolutionary origin of PB\,8 may be
derived from the planetary nebula. In Sect.\,\ref{sect:pb8-nebula} we
showed that the present nebula is younger than 3000\,years. There is no
visible remnant of an older PN. Moreover,
the nebula abundance ratios $\text{He}\,/\,\text{H}=0.123$ and
$\text{N}\,/\,\text{O}=0.28$ by number \citep{gar2009} show that PB\,8
is not a helium-enriched Peimbert's Type\,I PN. For the latter,
$\text{He}\,/\,\text{H}>0.125$ or $\text{N}\,/\,\text{O}>0.5$ is
expected \citep{peimbert1987}. A VLTP origin of the nebula is
therefore implausible. The low $\text{N}\,/\,\text{O}$ ratio also
indicates the absence of hot bottom burning (HBB), which is predicted
for more massive AGB stars. From a comparison with stellar
evolutionary tracks, \cite{kaler1989} deduce $\text{N}\,/\,\text{O} >
0.8$ as a sharp limit for N-enriched PNe, which are
supposed to indicate HBB in AGB-stars
with $M_\text{core} > 0.8\,\text{M}_\odot$.

Thus the following alternative scenarios might explain our results:
\begin{enumerate}
 \item The CSPN of PB\,8 has a low mass and evolves slowly. 
   For instance, a 
   $0.6\,\text{M}_\odot$ post-AGB star on the way to an LTP
    has a crossing time of $4000\,\text{years}$ from
   $10^4\,\text{K}$ to its maximum effective temperature
   \citep{bloecker2003}.
 Then, either
 \begin{enumerate}
  \item the present nebula was ejected by a born-again
    AGB-star after occurrence of a ``weak'' VLTP. A possible older PN
    from the first AGB phase has already dissolved. 
    As mentioned above, this scenario does not fit
    well to PB\,8, as the PN is not enriched in helium; or,
  \item the CSPN suffered an ``anomalous AFTP'', resulting in the
    observed surface abundances. The nebula was formed only during this
    AGB phase of the star. In this scenario it is
    difficult to explain the enhanced nitrogen abundance of the star,
    as nitrogen enrichment is neither predicted for the AFTP nor 
    can it originate from HBB.
 \end{enumerate}
 \item Alternatively, the CS of PB\,8 may have a relatively high mass
   and therefore may have evolved very fast. The crossing time
   for e.g.\ a $0.94\,\text{M}_\odot$ post-AGB star is only
   $50\,\text{years}$ \citep{bloecker2003}. A VLTP has already
   occurred, but most of the nebula observed now still
   originates from the first AGB period, not from the born-again AGB
   star after the VLTP. Albeit possible, this scenario has a low
   probability because the empirical mass distribution of central
   stars has a sharp maximum at $0.6\,\text{M}_\odot$ and declines
   substantially towards higher values
   \citep{tylenda2003}; furthermore there are no hints of
   HBB, which would be indicative for a more massive CSPN.
\end{enumerate}
However, one has to keep in mind that the appropriate
stellar evolution models are still calculated only in 1D. Especially
the convective mixing during H-ingestion flashes of the TPs were
treated in diffusion approximation, whereas recent hydrodynamical
studies, e.g.\ by \cite{woodward2008}, emphasize that convective
mixing is rather an advection process, making 2D or 3D calculations
necessary.

Alternatively to single star evolution, one may consider binarity with
a common envelope phase as the origin of the hydrogen
deficiency. However, PB\,8 shows no evidence of binarity. The nebula
does not look bipolar. 
Also, \cite{mendez1989} found no indication of radial
velocity variations between three spectra taken within one year, but
only changes of the P-Cygni line profiles, which must be attributed to
variability of the stellar wind. 

Summarizing, the evolutionary origin of PB\,8 cannot be explained by
any existing model for a post-AGB star which lost its hydrogen
envelope in a final thermal pulse. However, one can imagine
scenarios of a weak or anomalous thermal pulse, occurring on the
AGB or later, which may explain the unique chemical composition of
this star and its young nebula. 

The chemical composition found in the expanding atmosphere of the
central star of PB\,8 differs from any known central star
abundance. However, it resembles the rare transition class of WN/WC
subtypes of massive Wolf-Rayet stars. Therefore we suggest to open a
new class of [WN/WC]-type central stars with PB\,8 as its first
member. 
 
\acknowledgements
M. Pe\~{n}a acknowledges financial support from FONDAP-Chile and
DGAPA-UNAM (grants IN118405 and IN112708). This work was supported by
the Bundesministerium f\"ur Bildung und Forschung (BMBF) under grant
05AVIPB/1. 
M. Pe\~na is grateful to the Institute for Physics and Astronomy,
Potsdam University, for hospitality and financial support when 
part of this work was done.

\bibliographystyle{aa}

\bibliography{pb8.bib}

\end{document}